\documentclass[twocolumn,aps,prl,superscriptaddress,preprintnumbers,amsmath,amssymb,nofootinbib]{revtex4-2}
\usepackage{amsfonts}
\usepackage{mathrsfs}
\usepackage{bm}
\usepackage{braket}
\usepackage{dsfont}
\usepackage{slashed}
\usepackage{siunitx}
\usepackage{appendix}
\usepackage{physics}
\usepackage{commath}
\usepackage{graphicx}
\usepackage{float}
\usepackage[caption=false]{subfig}
\usepackage{booktabs}
\usepackage{array}
\usepackage{multirow}
\usepackage{microtype}
\usepackage{soul}
\usepackage[normalem]{ulem}
\usepackage{enumerate}
\usepackage[dvipsnames]{xcolor}

\definecolor{nicered}{rgb}{0.8,0.15,0.15}
\definecolor{linkblue}{HTML}{1F4E79}
\definecolor{bibcolor}{rgb}{0.2,0.2,0.7}

\usepackage[
    colorlinks=true,
    linkcolor=linkblue,
    citecolor=RoyalBlue,
    urlcolor=linkblue,
    bookmarks=true,
    hypertexnames=true
]{hyperref}

\usepackage{orcidlink}

\begin{document}

\title{Quantifying Information Hierarchy for Neutrino Oscillation Parameters at JUNO}

\author{Yu-han Shu}
\email{shuyh@ihep.ac.cn}
\affiliation{Institute of High Energy Physics, Chinese Academy of Sciences, Beijing 100049, China}
\affiliation{School of Physical Sciences, University of Chinese Academy of Sciences, Beijing 100049, China}

\author{Neetu Raj Singh Chundawat}
\email{chundawat@ihep.ac.cn}
\affiliation{Institute of High Energy Physics, Chinese Academy of Sciences, Beijing 100049, China}
\affiliation{Kaiping Neutrino Research Center, Guangdong 529386, China}

\author{Luis A. Delgadillo}
\email{ldelgadillof@ihep.ac.cn}
\affiliation{Institute of High Energy Physics, Chinese Academy of Sciences, Beijing 100049, China}
\affiliation{Kaiping Neutrino Research Center, Guangdong 529386, China}

\author{Yu-Feng Li}
\email{liyufeng@ihep.ac.cn}
\affiliation{Institute of High Energy Physics, Chinese Academy of Sciences, Beijing 100049, China}
\affiliation{School of Physical Sciences, University of Chinese Academy of Sciences, Beijing 100049, China}

\begin{abstract}
Since neutrinos are quantum systems inherently, the precision with which oscillation parameters can be estimated ultimately depends on how much information about these parameters is encoded in the neutrino state and how efficiently that information can be extracted through measurement. In this work, we quantify how information encoded in reactor antineutrino states flows through the measurement process to the events observed at the detector, using quantum and classical Fisher information. We establish the information ladder for JUNO, revealing that the loss of precision across different information levels is strongly parameter dependent. We demonstrate that the JUNO configuration approaches the optimal statistical limit for the oscillation parameters of the solar sector, while information on $\theta_{13}$ and $\Delta m_{31}^{2}$ is significantly degraded by the measurement strategy and detector effects. Despite this information loss, the remaining information is sufficient for JUNO to achieve sub-percent precision on $\Delta m_{31}^{2}$ within six years.
\end{abstract}

\maketitle
\newpage

\textit{\textbf{Introduction}---} Neutrino oscillations provide one of the most powerful probes of physics beyond the Standard Model, and achieving increasingly precise determination of the oscillation parameters has become a primary objective of current neutrino experiments. In the standard three-flavor oscillation picture, neutrino oscillations imply three non-degenerate mass eigenstates $\nu_i$ ($i=1,2,3$). Flavor eigenstates ($\nu_e,\nu_\mu,\nu_\tau$) can be written as superposition of mass eigenstates via the leptonic PMNS mixing matrix \cite{Pontecorvo:1957qd, Maki:1962mu, ParticleDataGroup:2024cfk}. The three flavor neutrino oscillations depends on six parameters: three mixing angles $\theta_{12}$, $\theta_{13}$, and $\theta_{23}$, one Dirac CP-violating phase $\delta_{\text{CP}}$, and two independent mass-squared differences $\Delta m^2_{ij}\equiv m_i^2-m_j^2$. 

Quantum estimation theory (QET) provides a framework for parameter estimation 
when the parameter of interest is encoded in a quantum state, and determines 
the ultimate precision with which it can be estimated through quantum 
measurements~\cite{Jaynes:1957zza,Rao:1945,Fisher:1925}. Different measurement strategies can be mathematically described 
by positive operator-valued measures (POVMs)~\cite{Nielsen:2010}. 
The ultimate precision achievable for a parameter, optimized over all possible 
POVMs, is characterized by the quantum Cramér-Rao bound (QCRB)~\cite{Helstrom:1969,Paris:2007xql,Pezze:2014dwa,Liu:2019xfr,Lu:2010ktl}. 
The QCRB is determined by the quantum Fisher information (QFI)~\cite{Braunstein:1994zz}, which depends 
solely on the parameterized quantum state and has important applications in quantum metrology~\cite{Giovannetti:2011chh,Albarelli:2020pec,Toth:2014msl}.

In recent studies in neutrino physics, QFI has emerged as a powerful tool for assessing the ultimate precision achievable in measurements of oscillation parameters at current and forthcoming neutrino experiments \cite{Nogueira:2016hta,Ignoti:2025rxr,Yadav:2026lsx,Frugiuele:2026yeq,Chundawat:2026jjd,Chundawat:2026lcm,Huang:2026bws,Yadav:2026mnw,Dixit:2026yft}. For instance, studies such as Refs.~\cite{Ignoti:2025rxr,Yadav:2026lsx,Frugiuele:2026yeq,Chundawat:2026lcm} demonstrate that QFI provides a systematic framework for evaluating the sensitivity of long-baseline (LBL) neutrino experiments to oscillation parameters, thereby identifying which parameters are most amenable to precise measurement. Likewise, in Ref.~\cite{Chundawat:2026jjd}, the authors explore information-theoretic difference in the estimation of solar oscillation parameters in solar and reactor neutrino experiments. Ref.~\cite{Chundawat:2026lcm} employs the QFI framework to investigate the information encoded in neutrino states and how efficiently it is extracted in current LBL experiments, including the impact of detector effects using reconstructed spectra. Additionally, multiparameter QFI analyses have recently been performed in Refs.~\cite{Huang:2026bws,Yadav:2026mnw}.

In this work, we develop a complete multiparameter quantum-estimation framework to trace oscillation-parameter information from its fundamental quantum limit, through the flavor measurement, to the information retained in the reconstructed event spectrum after including detector effects. Medium baseline reactor experiments provide an ideal setting to study this information hierarchy where coherence preserves the oscillation phase information, while finite energy resolution can smear the interference structure and further reduce the experimentally retained information.
For instance, the Jiangmen Underground Neutrino Observatory (JUNO) is expected to significantly improve the precision on $\Delta m^2_{31}$, $\Delta m^2_{21}$, and $\sin^2\theta_{12}$, while providing less competitive precision for the reactor mixing angle $\sin^2\theta_{13}$~\cite{JUNO:2022mxj}. However, it is insensitive to the remaining oscillation parameters, $\sin^2\theta_{23}$ and $\delta_{\text{CP}}$. The central objective of the JUNO experiment is to determine the neutrino mass ordering~\cite{Petcov:2001sy, Choubey:2003qx, Learned:2006wy, Zhan:2008id, Zhan:2009rs, Li:2013zyd, Bilenky:2017rzu, Forero:2021lax, Parke:2024xre, JUNO:2024jaw}. Recently, the JUNO experiment released its first results regarding detector performance~\cite{JUNO:2025fpc}, as well as the measurement of the solar mixing parameters $\Delta m^2_{21}$ and $\sin^2\theta_{12}$ with 59.1 days of data~\cite{JUNO:2025gmd}, improving the precision by a factor of \(1.6\) relative to the combination of previous measurements. Using 207 days of JUNO data, a preference for normal mass ordering at the $2.3\sigma$ level was also reported~\cite{Wang:2026juno}.
A summary of the current status of neutrino oscillation parameters can be found in Ref.~\cite{Esteban:2024eli}.

\textit{\textbf{Quantum-to-Classical Fisher Information}---}

In the three-dimensional flavor basis, an electron antineutrino state can be written as
\begin{equation}
    \ket{\bar\nu_e} =
    \begin{pmatrix}
        1 & 0 & 0
    \end{pmatrix}^{T}.
\end{equation}
The vacuum propagation of an initially produced electron antineutrino with energy $E_\nu$, assuming a real mixing matrix, is described by
\begin{equation}
    \ket{\psi({\lambda},L)}
    = UDU^{T}\ket{\bar\nu_e}.
    \label{eq:evolved-state}
\end{equation}
Here, we omit the $R_{23}(\theta_{23})$ rotation and define $U=R_{13}(\theta_{13})R_{12}(\theta_{12})$, since $\theta_{23}$ is not relevant to the present JUNO analysis. The parameter vector to be estimated is ${\lambda}=(\theta_{12},\theta_{13},\Delta m^2_{21},\Delta m^2_{31})$. Using the standard definitions, $\Delta_{21}\equiv\frac{\Delta m_{21}^{2}L}{4E_\nu}$, and $\Delta_{31}\equiv\frac{\Delta m_{31}^{2}L}{4E_\nu}$, the propagation matrix becomes
\begin{equation}
D=\operatorname{diag}\left(1,e^{-2 i\Delta_{21}},
e^{-2 i \Delta_{31}}\right).
\end{equation} 

For the pure quantum state $\ket{\psi({\lambda},L)}$, the multi-parameter QFI matrix is defined as,
\begin{align}
F^{Q}_{\alpha\beta}
=
4\,\operatorname{Re}\Big[
&\langle \partial_{\alpha}\psi({\lambda},L)
 | \partial_{\beta}\psi({\lambda},L)\rangle
\nonumber\\
&- \langle \partial_{\alpha}\psi({\lambda},L)
 | \psi({\lambda},L)\rangle
\langle \psi({\lambda},L)
 | \partial_{\beta}\psi({\lambda},L)\rangle
\Big]\;,
\label{eq:pureqfim}
\end{align}
where $\partial_{\alpha}\equiv\partial/\partial\lambda_{\alpha}$ and
$\partial_{\beta}\equiv\partial/\partial\lambda_{\beta}$ denote partial derivatives with respect to the parameters of interest. The ten independent elements of the QFI matrix are calculated and given in Appendix A.

An ideal measurement on a three flavor antineutrino state can be described by the projectors,
\begin{equation}
\Pi_x
=
\ket{\bar\nu_x}\bra{\bar\nu_x},
\qquad
x=e,\mu,\tau,
\qquad
\Pi_e+\Pi_\mu+\Pi_\tau=\mathbb{I}.
\label{eq:flavor-povm}
\end{equation}

For this complete flavor POVM, the classical Fisher information can be written as~\cite{Braunstein:1994zz}, 
\begin{equation}
F^{\rm flavor}_{\alpha\beta}(\lambda,E_\nu)
= \sum_{\gamma=e,\mu,\tau}
\frac{1}{P_{e\gamma}(\lambda, E_\nu)}
\frac{\partial P_{e\gamma}(\lambda,E_\nu)}{\partial\lambda_\alpha}
\frac{\partial P_{e\gamma}(\lambda,E_\nu)}{\partial\lambda_\beta}\;,
\label{eq:flavor-FI}
\end{equation}
where $\lambda_\alpha,\lambda_\beta\in \{\theta_{12},\theta_{13},\Delta m_{21}^{2},\Delta m_{31}^{2}\}$.

Reactor antineutrino experiments such as JUNO detect only the surviving electron antineutrinos through inverse beta decay (IBD) and do not separately measure the appearance channels. Therefore, the corresponding flavor Fisher information (flavor-FI) associated with the this channel is given by
\begin{equation}
F^{\bar e\bar e}_{\alpha\beta}(\lambda,E_\nu) = \frac{1}{P_{\bar e\bar e}(\lambda,E_\nu)} \frac{\partial P_{\bar e\bar e}(\lambda,E_\nu)}{\partial\lambda_\alpha} \frac{\partial P_{\bar e\bar e}(\lambda,E_\nu)}{\partial\lambda_\beta}\;,
\label{eq:electron-flavor-FI}
\end{equation}
further details about $P_{\bar e\bar e}(\lambda,E_\nu)$ and the ten independent elements of the electron flavor-FI matrix can be found in Appendix B.

At neutrino experiments, however, the reconstructed spectra is measured rather than the probability directly. 
We therefore demonstrate the construction of event-level Fisher information by considering an experimental configuration similar to JUNO~\cite{JUNO:2015zny,JUNO:2021vlw}. We employ the GLoBES software~\cite{Huber:2004ka,Huber:2007ji} to model a detector with JUNO-like characteristics. We consider a \SI{20}{kton} liquid scintillator detector located at \SI{52.5}{km} from the reactor core. Furthermore, we consider an exposure of $6~\text{years} \times 26.6~\text{GWth}$ of reactor thermal power. Besides, we employ the Huber-Mueller flux model~\cite{Huber:2011wv, Mueller:2011nm} to compute the predicted IBD signal events. Regarding backgrounds, we adopt the total expected background rates from Ref.~\cite{JUNO:2024jaw} including geoneutrinos, accidentals, $^9$Li/$^8$He, fast neutrons, world reactors, atmospheric neutrinos, and $^{13}$C($\alpha, n$)$^{16}$O contaminants. We consider constant binning with a bin size of $20$~keV for the reconstructed neutrino energy spectrum, spanning from $1.8$~MeV to $8$~MeV. Furthermore, the energy resolution function, which relates the true neutrino energy ($E_\nu$) to the reconstructed neutrino energy ($E_\nu^{\prime}$), follows a Gaussian distribution~\cite{Huber:2002mx}:
\begin{equation}
    R(E_\nu,E_\nu^{\prime}) =
    \frac{1}{\sqrt{2\pi}\,\sigma_{R}(E_\nu)}
    \exp\left[
    -\frac{(E_\nu-E_\nu^{\prime})^2}
    {2\sigma_{R}^2(E_\nu)}
    \right]\;,
\end{equation}
here the resolution is computed in terms of the prompt energy, $E_{\rm prompt}\simeq E_\nu-0.78~\si{MeV}$, such that
\begin{equation}
    \sigma_R(E_\nu) = \kappa \sqrt{\frac{E_\nu-0.78~\si{MeV}}{\si{MeV}}} \,\si{MeV}\;,
\end{equation}
with $\kappa=0.03$. This corresponds to an energy resolution of $3.0\%/\sqrt{E_{\rm prompt}/\si{MeV}}$ for the prompt energy related with the IBD signal. A similar experimental configuration was employed in Ref.~\cite{Delgadillo:2025wxw}.

The corresponding expected simulated events, including the IBD signal and backgrounds, is represented by
\begin{equation}
\mu_j^{\rm event}(\lambda) = \sum_k R_{jk}\, C_k\, P_{\bar e \bar e}(\lambda, E_k ) + B_j = S_j(\lambda) + B_j\;,
\end{equation}
where \(R_{jk}\) accounts for the energy response that maps events from the true-energy bin \(k\) to the reconstructed-energy bin \(j\). The term \(C_k\) represents the expected number of unoscillated antineutrino signal events in the \(k\)-th true-energy bin. Thus, \(S_j(\lambda)\) corresponds to the reconstructed antineutrino signal events after accounting for oscillations, while \(B_j\) represents the corresponding background events in the \(j\)-th reconstructed-energy bin, which are treated as fixed and independent of $\lambda$. Therefore, the event-level Fisher matrix can be constructed as~\cite{Chundawat:2026lcm}
\begin{equation}
\textbf{F}^{\rm event}_{\alpha \beta} = \sum_j \frac{1}{\mu_j^{\rm event}(\lambda)} \frac{\partial\mu_j^{\rm event}(\lambda)}{\partial\lambda_\alpha}
\frac{\partial\mu_j^{\rm event}(\lambda)}{\partial\lambda_\beta}\;.
\label{eq:Fevent}
\end{equation}
While the systematic uncertainties can be implemented through the covariance matrix or pull terms~\cite{Leane:2025efj}, we focus our analysis to the statistics-only case, which captures the essential sensitivity of a JUNO-like configuration.
\begin{figure*}[htbp]
\centering
\includegraphics[width=0.8\linewidth]{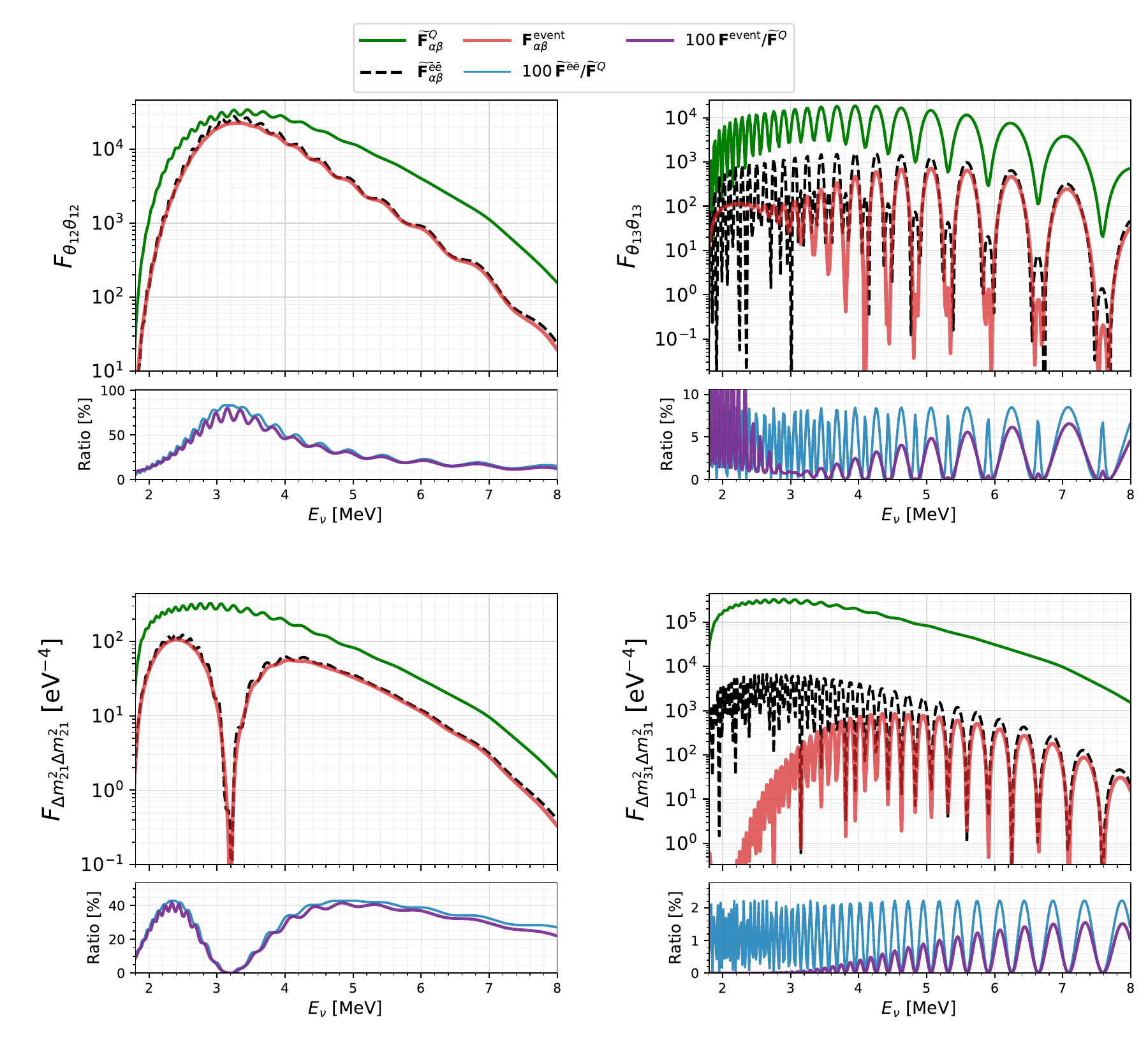}
\caption{
Diagonal Fisher-information elements for JUNO with $3\%$ energy resolution including signal and background contributions, shown for NO. Each subplot corresponds to one oscillation parameter: $\theta_{12}$, $\theta_{13}$, $\Delta m^2_{21}$, and $\Delta m^2_{31}$. The upper panels show the event-weighted QFI $\widetilde{\mathbf F}^{Q}_{\alpha\alpha}$, the event-weighted electron flavor-FI $\widetilde{\mathbf F}^{\bar e\bar e}_{\alpha\alpha}$, and the event-level Fisher information $\mathbf F^{\rm event}_{\alpha\alpha}$ as functions of the true antineutrino energy $E_\nu$. The lower panels show the corresponding information fractions, $\,\widetilde{\mathbf F}^{\bar e\bar e}_{\alpha\alpha}/\widetilde{\mathbf F}^{Q}_{\alpha\alpha}$(blue) and $\,\mathbf F^{\rm event}_{\alpha\alpha}/\widetilde{\mathbf F}^{Q}_{\alpha\alpha}$ (purple). }
\label{fig3:info_hierarchy}
\end{figure*}

For an ideal flavor measurement, characterized by perfect energy reconstruction, vanishing backgrounds, and an unoscillated spectrum independent of the oscillation parameters, the expected oscillation signal can be written as
\(S_k^{\rm ideal}(\lambda) =
C_k P_{\bar e\bar e}(\lambda,E_k)\;.\)
Thus, for a direct comparison with the event-level Fisher information defined in Eq.~\eqref{eq:Fevent}, the benchmark event-weighted electron flavor-FI matrix can be expressed in terms of the expected spectrum and the single electron flavor-FI matrix derived in the previous section as
\begin{equation}
\Tilde{\bf{F}}^{\bar e\bar e}_{\alpha \beta}
=
\sum_k C_k F^{\bar e\bar e}_{\alpha\beta}(E_k)\;.
\label{eq:event-flavor-limit}
\end{equation}
The difference between this benchmark event-weighted  electron flavor-FI and the event-level Fisher information represents the information loss caused by detector effects and backgrounds. Similarly, to compare these two Fisher information quantities given in Eqs.~\eqref{eq:Fevent} and \eqref{eq:event-flavor-limit} with the 
intrinsic quantum information, we construct the event-weighted QFI matrix as
\begin{equation}
\Tilde{\bf{F}}^Q_{\alpha \beta}
=
\sum_k C_k F_{\alpha\beta}^{Q}(E_k)\;.
\label{eq:weighted-qfim}
\end{equation}

\textit{\textbf{Information Loss and Parameter Precision---}} The hierarchy between the three information matrices defined in previous section can be understood as a quantum to classical information flow. This obeys 

\begin{equation}
    \widetilde{\mathbf F}^{Q}
    \succeq
    \widetilde{\mathbf F}^{\bar e\bar e}
    \succeq
    \mathbf F^{\rm event},
    \label{eq:information-hierarchy}
\end{equation}
where $\succeq$ denotes positive-semidefinite ordering.

The first inequality, between the QFI and the electron flavor-FI, follows from the fact that the QFI represents the largest Fisher information obtainable by optimizing over all possible POVMs, a standard result of QET. The flavor measurement at the detector selects a specific POVM and can therefore restrict the accessible information to the electron flavor-FI. The second inequality arises in accordance with the fundamental principles of information theory, from the further processing of this information by the detector response and backgrounds, resulting in the event-level Fisher information. While the hierarchy of the information matrices follows from general theoretical principles, the amount of information lost at each stage is not fixed and depends on both the oscillation parameter and the experimental setup. 

In Fig.~\ref{fig3:info_hierarchy}, we demonstrate how widely separated the corresponding diagonal Fisher information elements are for the oscillation parameters relevant to JUNO. A clear information hierarchy is observed over the entire energy range and for all four parameters, with the QFI~(green) providing the largest information, followed by the electron flavor-FI~(black), and finally the information retained in the event spectrum~(red). The gaps between these different information-theoretic quantities are strongly parameter dependent. For the solar mixing angle, the $\widetilde{\mathbf F}^{\bar e\bar e}_{\theta_{12}\theta_{12}}$  nearly saturates the $\widetilde{\mathbf F}^Q_{\theta_{12}\theta_{12}}$ around the peak of the JUNO neutrino energy spectrum. For $\Delta m_{21}^{2}$, the electron flavor-FI also remains close to the maximum information set by the QFI over almost the entire energy range, except around the peak-energy region. This demonstrates that the electron-antineutrino survival measurement provides a nearly optimal measurement strategy for estimating both solar oscillation parameters at JUNO. On the other hand, for both solar oscillation parameters the $\widetilde{\mathbf F}^{\bar e\bar e}$ and $\mathbf F^{\rm event}$ remain close to each other, indicating that the additional information loss due to detector effects at JUNO is mild.

Furthermore, for $\theta_{13}$, a considerable gap is observed between the QFI and the electron flavor-FI, indicating a substantial reduction of information at the flavor-measurement level. The event-level Fisher information shows an additional degradation at lower energies due to rapid oscillations and energy smearing, while a larger fraction of the flavor-accessible information is retained at higher energies. A similar behavior is observed for $\Delta m_{31}^{2}$, but with a more pronounced overall separation between the QFI and the event-level Fisher information.

The lower panels of Fig.~\ref{fig3:info_hierarchy} display the information-extraction efficiencies, expressed as the percentages of the electron flavor-FI and event-level Fisher information relative to their quantum counterpart. A substantial fraction of the available quantum information is extracted for the solar oscillation parameters, reaching maximum extraction efficiencies of around  $90\%$ for $\theta_{12}$ and $50\%$ for $\Delta m_{21}^{2}$. In contrast, the extraction efficiencies for $\theta_{13}$ and $\Delta m_{31}^{2}$ are considerably smaller, remaining below approximately $10\%$ and $2\%$, respectively.

To determine the attainable precision at each of the three information levels, we employ the Cramér-Rao bounds. For the ultimate quantum limit, independent of any specific measurement strategy, we use the QCRB, which states that the variance of any unbiased estimator of a parameter is bounded from below by the inverse of the product of the QFI and the number of independent measurements. Once the measurement strategy is specified for JUNO, the attainable precision is instead determined by its classical counterpart. In our analysis, we directly use the event-weighted Fisher information quantities to evaluate these bounds, since the expected event statistics, corresponding to the number of independent measurements, are already incorporated into their definition.

The diagonal or unmarginalized uncertainty on the parameter $\lambda_\alpha$ when all the other parameters are kept fixed is given by, 
\begin{equation}
\sigma_{\alpha}^{X,\mathrm{diag}}
\geq
\frac{1}{
\sqrt{F_{\alpha\alpha}^{X}}
}\,,
\end{equation}
where
\(
X\in
\left\{
Q,\,
\bar e\bar e,\,
\mathrm{event}
\right\}.
\)

For the simultaneous estimation of all the oscillation parameters relevant for JUNO, the correlations must be included by inverting the full Fisher information matrices. The marginalized uncertainty can be written as,
\begin{equation}
\sigma_{\alpha}^{X,\mathrm{marg}}
\geq
\sqrt{
\left[
\left(\bm F^{X}\right)^{-1}
\right]_{\alpha\alpha}
}\,.
\label{eq:general-crb}
\end{equation}
In general, the marginalization weakens the attainable precision, for each Fisher matrix we can write as
\begin{equation}
\sigma_{\alpha}^{X,\mathrm{marg}}
\geq
\sigma_{\alpha}^{X,\mathrm{diag}}\;,
\end{equation}
with equality holding for uncorrelated parameters.
\begin{figure*}
    \centering
    \includegraphics[width=0.8\linewidth]{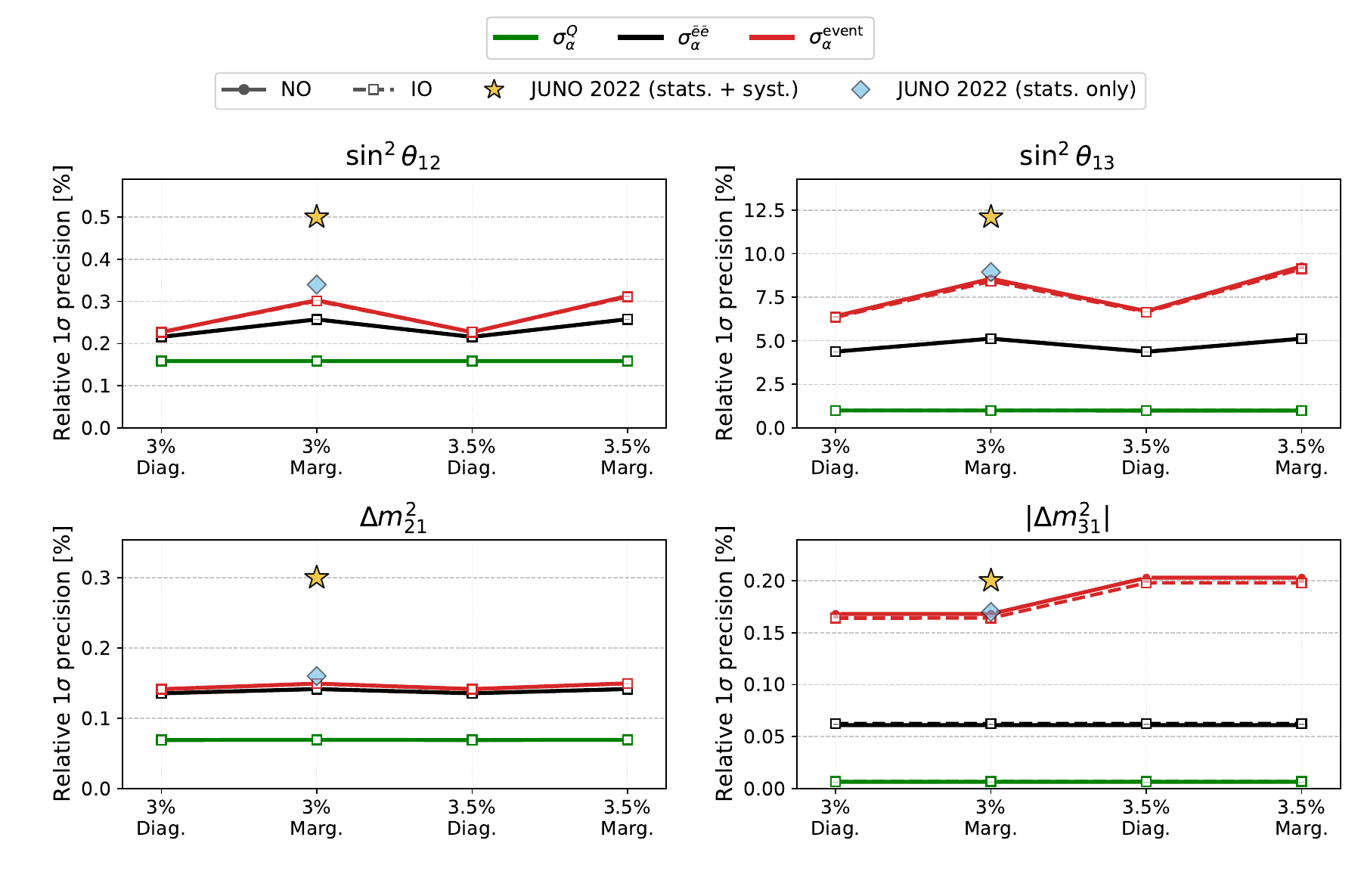}
    \caption{
Relative $1\sigma$ precision for the four oscillation parameters in JUNO, comparing diagonal and marginalized constraints for both NO and IO, as well as energy resolutions of $3\%$ and $3.5\%$. The diagonal values are extracted from the diagonal elements of the Fisher information matrix, while the marginalized values account for correlations via the inverse of the full Fisher matrix. Blue diamond markers indicate the statistics-only precision, whereas star markers denote the full statistics plus systematics projected six-year precision reported by the JUNO Collaboration~\cite{JUNO:2022mxj}.
}
    \label{fig:precision-comparison}
\end{figure*}

We present the relative precision obtained at all three levels of the Fisher-information hierarchy in Fig.~\ref{fig:precision-comparison}. The results are shown for both normal ordering (NO) and inverted ordering (IO), considering energy resolutions of $3\%$ and $3.5\%$ at $1~\mathrm{MeV}$. For comparison, we also include the benchmark precision values reported by the JUNO Collaboration~\cite{JUNO:2022mxj}. As expected, Fig.~\ref{fig:precision-comparison} shows a progressive loss of precision from the quantum-state level to the experimentally accessible event level for all four oscillation parameters, following the hierarchy \( \sigma^Q_\alpha \leq \sigma^{\bar e\bar e}_\alpha \leq \sigma^{\rm event}_\alpha.\)

For the two solar oscillation parameters, the gaps between these limits remain relatively small at both stages, from the quantum state to the electron-flavor measurement and subsequently from the flavor measurement to the detector level. In contrast, a considerably larger separation is observed for $\sin^2\theta_{13}$ and $\Delta m_{31}^{2}$ at the quantum-to-measurement level, indicating a substantial loss of information upon restricting the measurement to the electron-antineutrino survival channel. For $\Delta m_{31}^{2}$, an additional pronounced gap appears between the flavor-measurement and event levels, demonstrating its particular susceptibility to detector effects, especially energy smearing. This is further reflected in the reduction of this gap when the energy resolution is improved from $3.5\%$ to $3\%$. Overall, the precision hierarchy exhibits no significant dependence on the mass ordering, with nearly identical results for NO and IO. Marginalization predominantly affects the mixing-angle parameters, most notably $\sin^2\theta_{13}$, whereas its impact on the mass-squared splittings remains small.

\textit{\textbf{Conclusions}---}We quantify the systematic hierarchy of information in neutrino oscillations from neutrinos as the quantum state, through the flavor measurement strategy to the experimentally expected spectrum including detector effects. We demonstrate this reduction of information at each stage for the relevant oscillation parameters for JUNO. This information flow enables us to identify whether the sensitivity to a particular oscillation parameter is fundamentally limited by the quantum state, by the choice of measurement or by experimental effects. We find that, for JUNO, the estimation of the solar oscillation parameters is nearly optimal through the electron-flavor measurement and remains robust against detector effects and thus, efficiently translates information encoded in the neutrino state into experimentally accessible precision. Nevertheless, the information associated with $\theta_{13}$ and $\Delta m_{31}^{2}$ is strongly affected by the choice of measurement, with $\Delta m_{31}^{2}$ exhibiting an additional sensitivity to the finite detector energy resolution. Regardless of this information loss for $\Delta m_{31}^{2}$, JUNO retains sufficient information to achieve sub-percent precision on this parameter after six years of data taking. Our results provide an information-theoretic framework for interpreting JUNO's precision oscillation program by identifying where information is retained or lost from the quantum state to experimentally accessible events. More generally, this framework can be readily extended to other current and future neutrino oscillation experiments.

\textit{Acknowledgements---}We acknowledge Zeyuan Yu and J. Pedro Ochoa-Ricoux for their careful reading of the manuscript. NRSC\ and LAD\ were supported in part by the Kaiping Neutrino Research Center, China.

\begin{widetext}
\appendix

\section{Appendix A: QFI Matrix for JUNO}
\label{sec:qfim-juno}

\setcounter{equation}{0}
\renewcommand{\theequation}{A.\arabic{equation}}

The derivative of the evolved neutrino state given in Eq.~\eqref{eq:evolved-state} can be written as,
\begin{equation}
 \ket{\partial_\alpha\psi({\lambda},L)}
 = (\partial_\alpha U)DU^{T}\ket{\bar\nu_e}+U(\partial_\alpha D)U^{T}\ket{\bar\nu_e}.
 \label{eq:state-derivative}
\end{equation}
Multiplying the above equation by $U^{T}$ gives,
\begin{equation}
 \mathcal{A}_\alpha \equiv U^{T}\ket{\partial_\alpha\psi({\lambda},L)}
 =U^{T}(\partial_\alpha U)DU^{T}\ket{\bar\nu_e}+\partial_\alpha DU^{T}\ket{\bar\nu_e},
 \label{eq:A}
\end{equation}

The factor $U^{T}(\partial_\alpha U)$ accounts for the parameter-dependent rotation of the flavor basis and must therefore be included when calculating the QFI in that basis. Substituting Eq.~\eqref{eq:A} into the pure-state QFI expression in Eq.~\eqref{eq:pureqfim}, we obtain

\begin{equation}
 F_{\alpha\beta}^Q
 =4\operatorname{Re}\left[
 \mathcal{A}_\alpha^\dagger \mathcal{A}_\beta
 -(\mathcal{A}_\alpha^\dagger DU^{T}\ket{\bar\nu_e})((DU^{T}\ket{\bar\nu_e})^\dagger \mathcal{A}_\beta)
 \right].
 \label{eq:qfim-z}
\end{equation}

Since $F^{Q}_{\alpha\beta}=F^{Q}_{\beta\alpha}$, the QFI matrix is symmetric and has ten independent elements, which we calculate using Eq.~\eqref{eq:qfim-z}. The diagonal elements which quantifies the maximum information carried by a single antineutrino state about each oscillation parameter are given by,
\begin{equation}
F^{Q}_{\theta_{12}\theta_{12}}
=
16\cos^{2}\theta_{13}\sin^{2}\Delta_{21}
-
16\cos^{4}\theta_{13}\cos^{2}\theta_{12}\sin^{2}\theta_{12}
\sin^{2}(2\Delta_{21}),
\label{eq:F11}
\end{equation}
\begin{align}
F^{Q}_{\theta_{13}\theta_{13}}
&=
16\Bigg\{
\cos^{2}\theta_{12}\sin^{2}\Delta_{31}
+\sin^{2}\theta_{12}\sin^{2}(\Delta_{31}-\Delta_{21})
-\cos^{2}\theta_{13}\cos^{2}\theta_{12}\sin^{2}\theta_{12}
\sin^{2}\Delta_{21}
\nonumber\\
&\quad
-\cos^{2}\theta_{13}\sin^{2}\theta_{13}
\Big[
\cos^{2}\theta_{12}\sin(2\Delta_{31})
+\sin^{2}\theta_{12}
\sin\!\left(2(\Delta_{31}-\Delta_{21})\right)
\Big]^{2}
\Bigg\},
\label{eq:F22}
\end{align}
\begin{align}
F^{Q}_{\Delta m^{2}_{21}\Delta m^{2}_{21}}
&=
4\left(\frac{L}{2E_\nu}\right)^{2}
\cos^{2}\theta_{13}\sin^{2}\theta_{12}
\left(1-\cos^{2}\theta_{13}\sin^{2}\theta_{12}\right),
\label{eq:F33}
\end{align}
\begin{align}
 F^{Q}_{\Delta m^{2}_{31}\Delta m^{2}_{31}}
 &=4\left(\frac{L}{2E_\nu}\right)^{2}
 \sin^{2}\theta_{13}\cos^{2}\theta_{13}.
 \label{eq:F44}
\end{align}

For $\theta_{12}$, the information only depends on the solar oscillation phase, whereas for $\theta_{13}$ it depends on both atmospheric phases ($\Delta_{31}, \Delta_{31}-\Delta_{21}$) and the solar oscillation phase ($\Delta_{21}$). For mass-squared differences, the information scales as $(L/E_\nu)^2$ and, is suppressed by $\sin^{2}\theta_{13}$ for $\Delta m^{2}_{31}$ relative to the QFI for $\Delta m^{2}_{21}$. The off-diagonal element for the two mixing angles is given by,
\begin{align}
F^{Q}_{\theta_{12}\theta_{13}}
&=
8\cos\theta_{13}\sin\theta_{13}
\cos\theta_{12}\sin\theta_{12}
\Bigg\{
\cos\!\left[2(\Delta_{31}-\Delta_{21})\right]
-\cos(2\Delta_{31})
\nonumber\\
&\quad
-2\cos^{2}\theta_{13}\sin(2\Delta_{21})
\Big[
\cos^{2}\theta_{12}\sin(2\Delta_{31})
+\sin^{2}\theta_{12}
\sin\!\left[2(\Delta_{31}-\Delta_{21})\right]
\Big]
\Bigg\}.
\label{eq:F12}
\end{align}

To connect the two mass-squared differences, the off-diagonal element can be written as, 
\begin{align}
 F^{Q}_{\Delta m^{2}_{21},\Delta m^{2}_{31}}
 &=-4\left(\frac{L}{2E_\nu}\right)^{2}
\cos^{2}\theta_{13}\sin^{2}\theta_{12}\sin^{2}\theta_{13}.
 \label{eq:F34}
\end{align}
The correlation of information between the two solar oscillation parameters is given by,
\begin{align}
F^{Q}_{\theta_{12}\Delta m^{2}_{21}}
&=
4\left(\frac{L}{2E_\nu}\right)
\cos^{2}\theta_{13}\cos\theta_{12}\sin\theta_{12}
\sin(2\Delta_{21})
\left(1-2\cos^{2}\theta_{13}\sin^{2}\theta_{12}\right).
\label{eq:F13}
\end{align}

The off-diagonal elements relating the mixing angle and the mass-squared difference of other oscillation sector are,
\begin{align}
F^{Q}_{\theta_{12}\Delta m^{2}_{31}}
&=
-8\left(\frac{L}{2E_\nu}\right)
\cos^{2}\theta_{13}\cos\theta_{12}\sin\theta_{12}
\sin^{2}\theta_{13}
\sin(2\Delta_{21})
\label{eq:F14}
\end{align}
and 
\begin{align}
F^{Q}_{\theta_{13}\Delta m^{2}_{21}}
&=
4\left(\frac{L}{2E_\nu}\right)
\cos\theta_{13}\sin\theta_{13}\sin^{2}\theta_{12}
\Bigg\{
\sin\!\left[2(\Delta_{31}-\Delta_{21})\right]
\nonumber\\
&\quad
-2\cos^{2}\theta_{13}
\Big[
\cos^{2}\theta_{12}\sin(2\Delta_{31})
+\sin^{2}\theta_{12}
\sin\!\left[2(\Delta_{31}-\Delta_{21})\right]
\Big]
\Bigg\},
\label{eq:F23}
\end{align}
whereas the correlation between $\theta_{13}$ and $\Delta m^{2}_{31}$ is given by,
\begin{align}
F^{Q}_{\theta_{13}\Delta m^{2}_{31}}
&=
4\left(\frac{L}{2E_\nu}\right)
\cos\theta_{13}\sin\theta_{13}
\left(1-2\sin^{2}\theta_{13}\right)
\Big[
\cos^{2}\theta_{12}\sin(2\Delta_{31})
+
\sin^{2}\theta_{12}
\sin\!\left[2(\Delta_{31}-\Delta_{21})\right]
\Big].
\label{eq:F24}
\end{align}

At $L=0$, all elements of the QFI matrix vanish because the initial flavor state does not depend on any oscillation parameter. Since $[L/(2E_\nu)]=(\mathrm{eV}^{2})^{-1}$ in natural units, the matrix elements involving one mass-squared difference have dimensions $[F^{Q}_{\theta,\Delta m^{2}}]=(\mathrm{eV}^{2})^{-1}$, while those involving two mass-squared differences have dimensions $[F^{Q}_{\Delta m^{2},\Delta m^{2}}]=(\mathrm{eV}^{2})^{-2}$. The mixing-angle elements are dimensionless, $[F^{Q}_{\theta\theta}]=1$. 

\begin{figure}[H]
    \centering
    \includegraphics[width=1\linewidth]
    {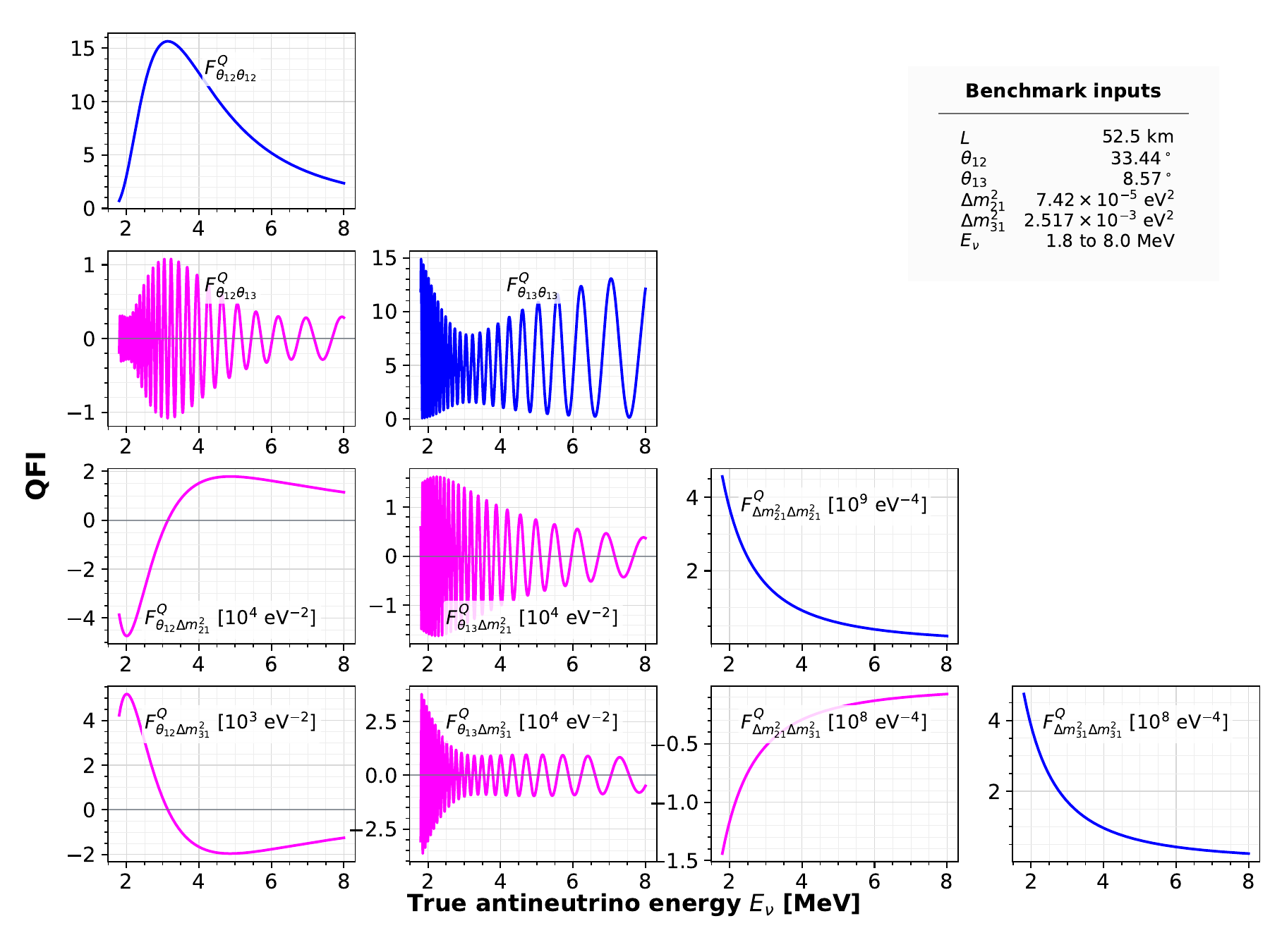}
    \caption{Energy dependence of the single-neutrino QFI matrix elements for an electron antineutrino propagating
    in vacuum at the JUNO baseline.   }
    \label{fig:single-qfim}
\end{figure}

The energy dependence of all QFI matrix elements at the JUNO baseline is shown in Fig.~\ref{fig:single-qfim}. The diagonal elements quantify the information available for estimating each parameter, whereas the off-diagonal elements describe their joint dependence. As evident from the figure, all diagonal elements are positive, while the off-diagonal elements may be either positive or negative, provided that the full QFI matrix remains positive semidefinite. The QFI element for $\theta_{12}$ is governed by $\Delta_{21}$ and exhibits a maximum at neutrino energies of approximately $3$--$4~\mathrm{MeV}$. The element $F^{Q}_{\theta_{13}\theta_{13}}$ shows rapid oscillations due to the atmospheric phases. Both mass-squared diagonal elements decrease gradually with increasing energy because of their $(L/E_\nu)^2$ dependence. The off-diagonal QFI matrix elements describe how variations in the oscillation parameters are encoded in the neutrino state. Elements involving $\theta_{13}$ exhibit rapid oscillations, while the others vary more smoothly. The element $F^{Q}_{\Delta m^{2}_{21}\Delta m^{2}_{31}}$ remains negative throughout the JUNO energy range.

\section{Appendix B: Flavor-FI matrix for JUNO}
\label{sec:flavorfi-juno}

\setcounter{equation}{0}
\renewcommand{\theequation}{B.\arabic{equation}}

To estimate the electron flavor-FI given in above equation, we use the electron-antineutrino survival probability in vacuum~\cite{JUNO:2022mxj}  as given by, 

\begin{equation}
P_{\bar e\bar e}(\lambda,E_\nu)
=
1
-
\sin^{2}2\theta_{12}\cos^{4}\theta_{13}\sin^{2}\Delta_{21}
-
\sin^{2}2\theta_{13}
\left[
\cos^{2}\theta_{12}\sin^{2}\Delta_{31}
+
\sin^{2}\theta_{12}\sin^{2}(\Delta_{31}-\Delta_{21})
\right].
\label{eq:prob_vac}
\end{equation}

\begin{figure}[H]
\centering
\includegraphics[width=1\linewidth]{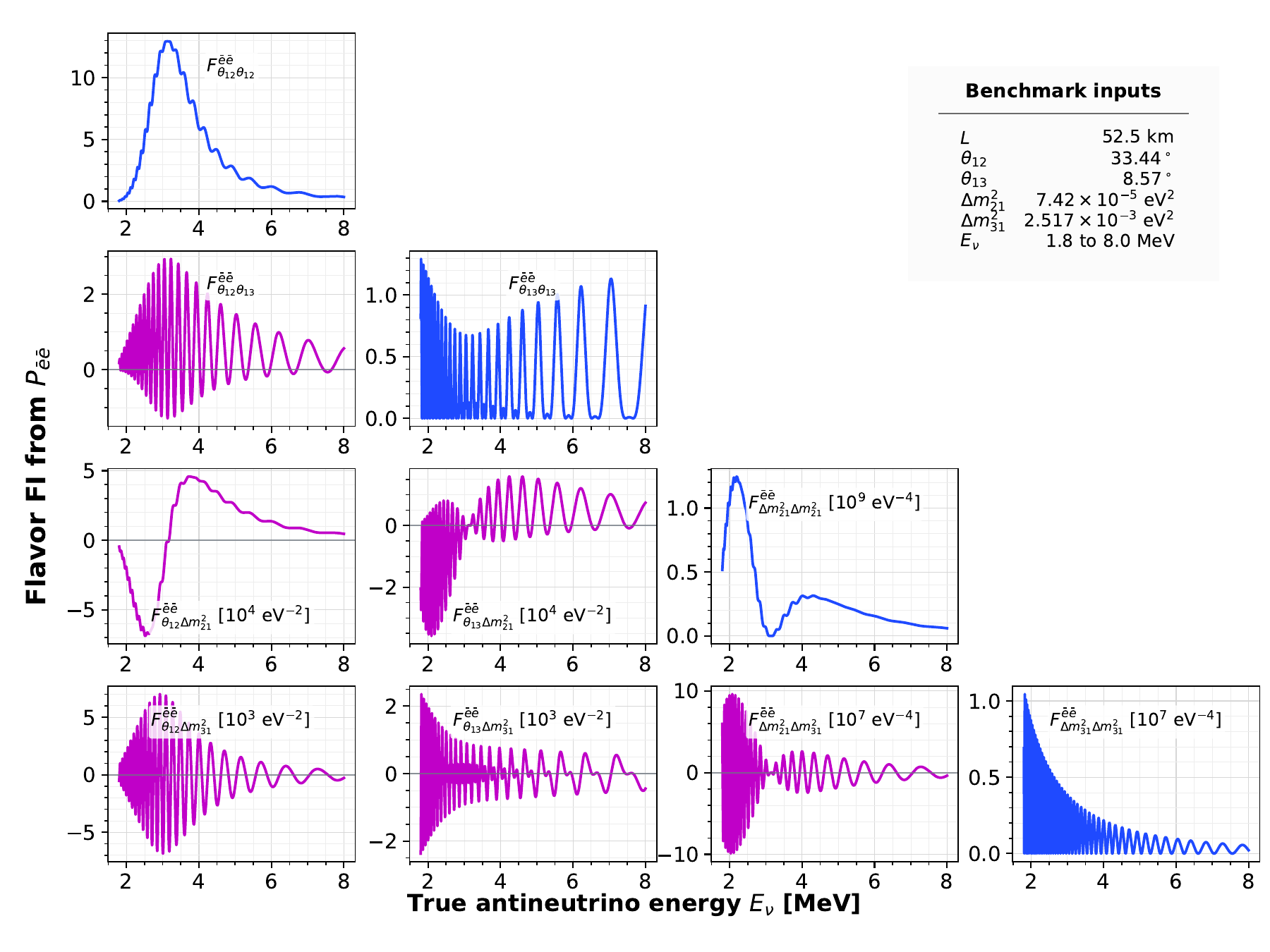}
\caption{Energy dependence of the ten independent elements of the electron-flavor Fisher information matrix obtained from the survival probability $P_{\bar e \bar e}$ for the four oscillation parameters considered in the JUNO analysis.}
\label{fig:flavor-fi}
\end{figure}

In the present analysis, matter effects are neglected. This approximation can produce biases of approximately $1\%$ and $0.2\%$ in the determination of $\Delta m^{2}_{21}$ and $\sin^{2}\theta_{12}$, respectively~\cite{Li:2016txk}. Detailed studies of matter effects in JUNO can be found in Refs.~\cite{Li:2016txk,Capozzi:2013psa,Khan:2019doq}. The electron-flavor Fisher information characterizes the information about the oscillation parameters accessible through the survival-channel measurement. 

Comparing this matrix with the QFI matrix derived in Appendix A, indicates how much of the information encoded in the propagated antineutrino state can be extracted through electron-flavor detection. Similar to the QFI matrix, we can derive all ten independent elements corresponding to the electron flavor-FI. The four diagonal elements quantifies the sensitivity of $P_{\bar e\bar e}(\lambda,E_\nu)$ to the variations in each oscillation parameter. The six off-diagonal elements represent the dependence of $P_{\bar e\bar e}(\lambda,E_\nu)$ on set of two parameters.

In Figure~\ref{fig:flavor-fi}, we show the ten independent elements of the electron flavor-FI matrix as functions of the true neutrino energy. The information associated with the solar and atmospheric oscillation scales exhibits distinct behavior. The diagonal elements corresponding to the solar oscillation parameters contain more information in the low-energy region, whereas the elements involving $\theta_{13}$ and $\Delta m_{31}^{2}$ exhibit rapid oscillations over the entire energy range. The positive and negative contributions of the off-diagonal elements are also visible in their energy dependence. Thus, the figure demonstrates that the different oscillation parameters possess distinct information signatures in the electron flavor-FI matrix.

\end{widetext}

\end{document}